\documentclass[preprint]{article}

\usepackage[nonatbib]{neurips_2021}
\usepackage{amsmath}
\usepackage{amsfonts}
\usepackage{graphicx}
\usepackage{caption}
\usepackage{subcaption}

\usepackage[square, numbers]{natbib}
\usepackage[utf8]{inputenc} 
\usepackage[T1]{fontenc}    
\usepackage{hyperref}       
\usepackage{url}            
\usepackage{booktabs}       
\usepackage{amsfonts}       
\usepackage{nicefrac}       
\usepackage{microtype}      
\usepackage{xcolor}         

\DeclareMathOperator*{\argmin}{arg\,min}
\newcommand{\Eqnref}[1]{Eq.~\ref{eq:#1}}
\newcommand{\Eqsref}[1]{Eqs.~\ref{eq:#1}}
\newcommand{\Figref}[1]{Fig.~\ref{fig:#1}}

\newcommand{\naug}{\ensuremath{n_\mathrm{aug}}\,}
\newcommand{\nee}{\ensuremath{n_\mathrm{e}}\,}

\title{Machine learning accelerated particle-in-cell plasma simulations}

\author{%
  R. Kube 
    \\
  Princeton Plasma Physics Laboratory\\
  Princeton, NJ 08540 USA \\
  \texttt{rkube@pppl.gov} \\
  \And
  R.M. Churchill \\
  Princeton Plasma Physics Laboratory\\
  Princeton, NJ 08540 USA \\
  \texttt{rchurchi@pppl.gov} \\
  \AND
  B. Sturdevant \\
  Princeton Plasma Physics Laboratory \\
  Princeton, NJ 08540 USA \\
  \texttt{bsturdev@pppl.gov} \\
}

\begin{document}

\maketitle

\begin{abstract}
  Particle-In-Cell (PIC) methods are frequently used for kinetic, high-fidelity simulations of plasmas.
  Implicit formulations of PIC algorithms feature strong conservation properties, up to numerical round-off errors,
  and are not subject to time-step limitations which make them an attractive candidate to use in simulations 
  fusion plasmas. Currently they remain prohibitively expensive for high-fidelity simulation of macroscopic plasmas. 
  We investigate how amortized solvers can be incorporated with PIC methods for simulations of plasmas. Incorporated into the
  amortized solver, a neural network  predicts a vector space that entails an approximate solution of the PIC system.
  The network uses only fluid moments and the electric field as input and its output is used to augment the vector space
  of an iterative linear solver. We find that this approach reduces the average number of
  required solver iterations by about $25\%$  when simulating electron plasma oscillations. This novel approach may
  allow to accelerate implicit PIC simulations while retaining all conservation laws and may also be appropriate for multi-scale systems.
\end{abstract}

\section{Background}

The dynamics of plasmas are governed by the Vlasov-Maxwell equations, which describe the self-consistent time-evolution of the plasmas
distribution function in phase space and the electromagnetic field. While approximations can be made to derive reduced
fluid models from this equation set, many situations require to resolve kinetic effects in simulations. Particle-in-cell (PIC) algorithms
power predictive high-fidelity kinetic simulations of fusion plasmas, implemented for example in the XGC code \cite{ku-2016}. Such
simulations resolve complicated multi-scale interactions on a kinetic level in order to uncover the physics that drive turbulent transport
in nuclear fusion experiments \cite{chang-2017, ku-2018, hager-2020}. 
Simulations of macroscopic systems which employ kinetic formulations of the physics model quickly become prohibitively expensive, 
even for leadership class, pre-exascale compute facilities. Researchers are therefore constantly implementing advanced parallelization
and optimization methods, often tailored to specific target hardware architectures, in order to speed up such codes
\cite{wang-2019, chien-2020, bird-2021, ohana-2021, Mniszewski-2021}.
In this contribution we are reporting on the development of a machine learning aided amortized solver that accelerates PIC simulations.
In particular, we focus on the method described by \citet{chen-2011, chen-2014}, which has been adapted for the XGC code to study 
electromagnetic phenomena in fusion plasmas \cite{sturdevant-2021}.

PIC algorithms couple the evolution of plasma distribution function and electromagnetic fields in a self-consistent manner.
An ensemble of marker particles represents the plasmas distribution function. Their coordinates typically assume continous
values, i.e. a Lagrangian description is used. The electromagnetic field on the other hand is commonly discretized in a spatial grid.
Evolving this system in time happens is a four-step process. 
First, the particles are pushed by integrating their equations of motion. Second, the electric current and mass densities
are evaluated on the grid, using the updated particle coordinates. Third, the electromagnetic field is updated using
the new source terms. And finally, electromagnetic forces on the individual particles are calculated from the updated field.
Implicit formulations of this algorithm can be formulated as to conserve energy and charge up to numerical round-off errors
\cite{chen-2011, chen-2014}. 


Time-stepping of implicit PIC algorithms requires to solve a system of non-linear equations. Jacobian-free Newton Krylov methods,
such as GMRES, are known to effectively solve such systems \cite{chen-2011}. However, to evaluate the effect of the systems Jacobian
acting on a vector requires to evaluate the systems time-stepping loop. This is the most expensive part of implicit PIC algorithms
and it is desirable to minimize the number of GMRES iterations required to integrate the system in time.  
In this contribution we explore how machine learning can be used to minimize the number of GMRES iterations required to
advance the PIC system in time. In particular, Neural Networks are trained to suggest vectors which augment the initial Krylov space of GMRES iterations
\cite{morgan-1995, chapman-1997, gaul-2014}. Interfacing a predictive model with a robust iterative solver has several
advantages when aiming to accelerate PIC simulations of plasmas. 
First, the simulation still obeys all conservation laws inherent to the used discretization. Meeting this condition is crucial to ensure
physically correct simulations. Second, by predicting a set of augmentation vectors, the model may predict only values of order unity. 
PIC algorithms are used to solve multi-scale problems where no single characteristic scale can be defined for the physical quantities in the system. 
If a model were to predict physical quantities it would eventually be required to make predictions over multiple orders of magnitude,
which is a hard problem.

Other contemporary approaches that aim to accelerate numerical simulations mostly target systems that are described by a set of partial
differential equations on an eulerian grid, such as the Navier-Stokes fluid equations. One thread of research aims to develop surrogate models, 
for example physics-informed neural networks \cite{raissi-2019} while other approaches use machine learned sub-grid models to accelerate
simulations \cite{pathak-2020, kochkov-2021}. Properly trained, such models allow to run coarse-grid simulations that include the same
level of detail as fine-grid models and result in speed-ups up to 100x. However, the hybrid Lagrangian-Eulerian structure of PIC methods
is not readily amenable to such approaches. 

\section{Interfacing an implicit PIC solver and predictive machine learning models}

Considering a collisionless, electrostatic plasma, an implicit discretization of the Vlasov-Ampere system can be 
formulated as
\begin{subequations}
\begin{align}
    \frac{x_\mathrm{p}^{n+1} - x_\mathrm{p}^{n}} {\triangle t} & = v_\mathrm{p}^{n+1/2} \label{eq:PIC_timestep_x}\\
    \frac{v_\mathrm{p}^{n+1} - v_\mathrm{p}^{n}}{\triangle t}  & = \frac{q_\mathrm{p}}{m_\mathrm{p}} \mathrm{SM}\left[ E^{n+1/2} \right] \left( x_\mathrm{p}^{n+1/2} \right) \label{eq:PIC_timestep_v}\\
    \epsilon_0 \frac{E^{n+1}_{i} - E^{n}_{i}}{\triangle t} + \frac{q_\mathrm{p}}{m_\mathrm{p}} \mathrm{SM}\left[ j_{i}^{n+1/2} \right] & = \langle j \rangle^{n+1/2} \label{eq:PIC_timestep_E}.
    \end{align}
    \label{eq:PIC_timestep}
\end{subequations}
Here $x_\mathrm{p}$ and $v_\mathrm{p}$ denote position and velocity of particle $\mathrm{p}$, 
$\triangle t$ denotes the time step length and 
$q_\mathrm{p}$ and $m_\mathrm{p}$ denote the particles charge and mass respectively. 
Superscript indices denote the time step and half-step quantities are given by an arithmetic mean, $x^{n+1/2} = \left(x^{n+1} + x^{n} \right) / 2$. 
The electric field $E$ is discretized on a periodic domain $z_i = i\, \triangle z \in L_z$, where $i=0\ldots N_z-1$ and $\triangle z = L_z / N_z$. The vaccuum permittivity is denoted as $\epsilon_0$ and 
$\mathrm{SM}[\cdot]$ denotes a binomial smoothing operator. The electric current is denoted as $j$ and $\langle \cdot \rangle$ denotes
spatial averaging. 

Given particle coordinates and the electric field at time step $n$, $x^{n}$, $v^{n}$, $E^{n}$, \Eqsref{PIC_timestep} guides the time
evolution of the system. Only for the physically correct $x^{n+1}$, $v^{n+1}$ and $E^{n+1}$ do these equations hold.
Re-casting \Eqsref{PIC_timestep} as a set of non-linear equations $G(E(\{x_\mathrm{p}\},\{v_\mathrm{p}\})) = 0$ allows to solve
this system for the correct values at $t= (n+1) \triangle t$ using Newton's method. That approach requires to solve the linear system of equations
\begin{align}
    \left. \frac{\partial G}{\partial E} \right|^{k} \delta E^{k} & = -G(E^{k}) \label{eq:Newton_Axb}
\end{align}
at each Newton iteration $k$. Here we denote the iterative solution as $\delta E^{k} = E^{k+1} - E^{k}$. 
Probing Jacobian-vector products $\partial G / \partial E |^{k} v$ by evaluating Gateaux derivative, \Eqnref{Newton_Axb} is solved
using GMRES, an iterative Krylov subspace method.
Each evaluation of  $\partial G / \partial E |^{k} v$ evaluates the entire four-step PIC loop. It is thus desirable to reduce the number
of iterations to converge the residual below the level of tolerance.

Neglecting preconditioning, the number of GMRES iterations required to solve a system $Ax=b$ within a prescribed residual tolerance may be reduced
by starting with a good initial guess. We therefore train a machine learning model to suggest vectors $\left\{  v_1,  \ldots, v_{\naug} \right\}$
such that an initial guess $A^{-1}b$ lies closely in $\mathcal{V}$. In particular, the solution vector $x$ is decomposed as
\begin{align}
    x = \sum_{j=1}^{\naug} \alpha_j v_j + \widetilde{x},
\end{align}
where $x_v = \sum_j \alpha_j v_j \in \mathcal{V}$ and $\widetilde{x} \perp \mathcal{V}$ and 
$\mathcal{V} = \mathrm{span} \left( v_1,  \ldots, v_{\naug}  \right)$. The coefficients $\alpha_j$ are
recovered by inserting the decomposition $x = x_v + \widetilde{x}$ into $Ax=b$, forming the scalar products between
this equation and the vectors $\left\{ Av_j \right\}_{j=1}^{\naug}$, and then solving the resulting system for $\alpha_j$.
Forming $\widetilde{b} = b - \sum_{j=1}^{\naug} \alpha_j A v_j$ we then proceed to solve
\begin{align}
    A \widetilde{x} = \widetilde{b} \label{eq:modified_system}
\end{align}
and reconstruct $x = \widetilde{x} + \sum_{j=1}^{\naug} \alpha_j A v_j$. This projection into
a space perpendicular to $\mathcal{V}$, the solution of \Eqnref{modified_system}, and the subsequent back-projection taken together 
constitute the amortized solver.




\begin{figure}
\begin{subfigure}[b]{0.45\textwidth}
    \includegraphics[width=\textwidth]{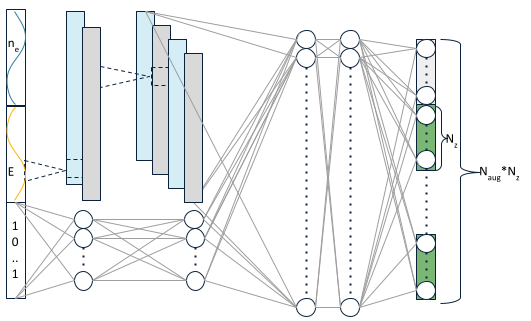}
    \caption{Convolutional neural network architecture used in the amortized solver that maps simulation profiles onto a set of 
    $\naug$ augmentation vectors $v_1, \ldots\, v_{\naug}$.}
    \label{fig:nnv6_architecture}
\end{subfigure}
\hfill
\begin{subfigure}[b]{0.45\textwidth}
    \includegraphics[width=\textwidth]{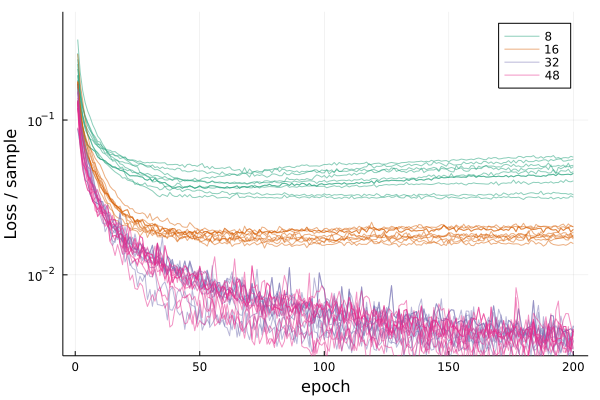}
    \caption{Loss over the test set, grouped by the number of augmentation vectors $\naug$ returned by the model.}
    \label{fig:test_loss}
\end{subfigure}
\caption{Network architecture (left) and test losses (right)}
\end{figure}

A convolutional neural network, shown  in \Figref{nnv6_architecture}, maps the electron density profile $\nee(z)$, 
the Electric field $E(z)$ profile and a one-hot encoding of simulation parameters onto $\left\{ v_j \right\}_{j=1}^{\naug}$. The profiles are concatenated and feed into
two convolutional layers with $4$ and $16$ channels respectively. The one-hot encoded simulation parameters feed into
two fully-connected MLP layers with $32$ neurons each. The output of these channels are flattened and merge into two
fully-connected layers which are fed into the output layer of size $N_z \times \naug$.
We train this network on data from a series of simulations where the initial positions of $32,768$ electrons is sampled
from profiles of the form $\nee(z, t=0) = A \cdot f(k z)$ with 
$A \in \{1, 2, 5, 10\} \times 10^{-3}$, $f \in \left\{ \sin, \cos \right\}$, $k \in \{1, 2\}$. The initial positions of $32,768$ singly
charged ions is sampled from a uniform distributions and both electrons and ions are initially cold. These initial conditions
on the particles are chosen as to
excite electron plasma oscillations. The fields are evaluated on a grid with $N_z=64$ and $L_z = 4\pi$. These 16 reference 
simulations are advanced with a timestep of $\triangle t=1$ from $T=0$ to $T=30$. This yields $480$ training profiles.

Targeting only the first Newton iteration, the neural network is trained by minimizing the orthogonal projection of
$\left\{ \delta E^{k=1} \right\}$ over the vector space spanned by the model output, as described by \citet{trivedi-2019}
\begin{align}
    \mathcal{L} = \argmin_{\widehat{y} \in \mathrm{span}(v_1\, \ldots, v_{\naug})} \frac{\| \widehat{y} - \delta E^{k=1} \|_2 }{\| \delta E^{k=1} \|_2} + \lambda \, \| \mathrm{triu} \left( V*V^{\mathrm{tr}} \right) \|_2. \label{eq:lossfun}
\end{align}
To calculate the projection, the output of the model $V$ is subject to a $\mathrm{QR}$ factorization, $Q,R = \mathrm{qr}\left(V\right)$
\cite{seeger-2017} and the projection is calculated as $\widehat{y} = Q \cdot Q' \cdot \delta E^{k=1}$.
Additionally, the last term penalizes collinearity of output vectors. Here $\mathrm{triu}(A) = \left\{ a_{r,s} \right\}_{r<s}$
denotes the upper triangular part of a square matrix and $V = \left[ v_1 | \ldots | v_{\naug}\right]$.


To identify optimal hyperparameters of the network, the $480$ profiles are split $80\%/20\%$ into a training set and a test set.
The network is trained for 200 epochs to minimize \Eqnref{lossfun} over the training set. We use
ReLU, tanh, and swish \cite{ramachandran-2017} activation functions, 
vary the size of the convolution filters used in both convolutional layers, choosing from $\{ 3, 5, 7, 9, 13\}$, 
vary dropout probabilities in fully connected layers \cite{srivastava-2014} over $\{0.1, 0.2, 0.3\}$,
vary $\lambda \in \{10^{-3}, 10^{-4}, 10^{-5}\}$
vary the initial learning rate from $\{ 10^{-2}, 10^{-3}, 5 \times 10^{-4}\}$, 
and use ADAM \cite{kingma-2017-adam}, RMSProp and SGD optimizers. 
Average loss over the test set serves as the metric to evaluate optimal hyperparameters. 
Figure \ref{fig:test_loss} shows a sub-set of the training losses for various $\naug$, $\lambda$ and dropout probabilities.
We observe that more augmentation vectors results in lower per-sample loss. Additionally we observe only
little over-fitting for $\naug = 8, 16$. From this scan we identify 
ReLU activation functions, 
convolutional filters of width $3$, 
a Dropout probability of $0.2$, 
$\lambda=10^{-4}$,
as ideal parameters 
and optimize using the ADAM algorithm with an initial learning rate of $0.001$.
Training was performed on nVidia V100 GPUs, in less than 2 hours of total computation time, using the Flux library \cite{Flux.jl-2018, innes-2018}. 
Simulation data for training and inference was produced using this code \footnote{\url{https://github.com/rkube/picfun}}. Instructions
on how to generate the data are available from the author upon request.

\begin{figure}[h!tb]
\begin{subfigure}[b]{0.49\textwidth}
    \includegraphics[width=\textwidth]{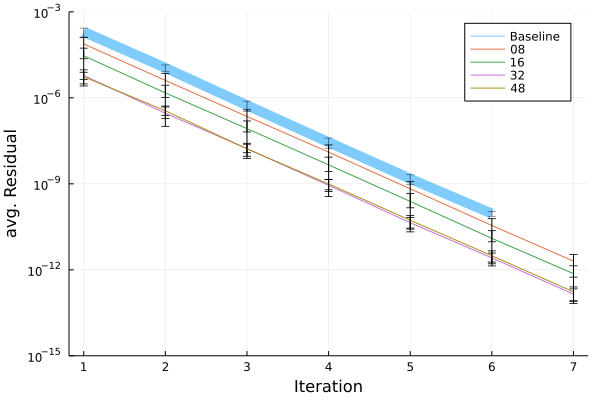}
    \caption{$n_{\mathrm{e},t=0}(z)=10^{-2} \left[ 0.1 \cos(2z) + 0.9 \sin(z)\right]$.}
    \label{fig:avgres_case1}
\end{subfigure}
\hfill
\begin{subfigure}[b]{0.49\textwidth}
    \includegraphics[width=\textwidth]{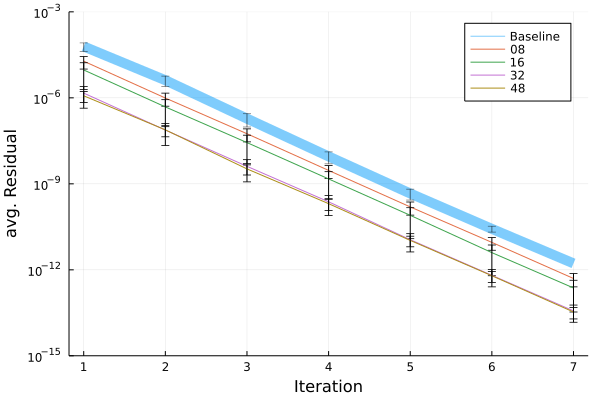}
    \caption{$n_{\mathrm{e},t=0}(z)=3\times10^{-3} \left[ 0.5 \sin(z) + 0.5 \cos(2z)\right]$.}
    \label{fig:avgres_case2}
\end{subfigure}
\caption{Residuals at various GMRES iterations of the original and amortized solver using $\naug \in \left\{ 8, 16, 32, 48\right\}$,
         averaged over an entire simulation.}
\end{figure}

The performance of the trained amortized solver is evaluated on simulations where the initial 
positions of the electrons are sampled from 
$n_{\mathrm{e}, t=0}(z) = A\left[ a_1 f_1(k_1 z) + a_2 f_2 (k_2 z)\right]$. Ions are again sampled from
a uniform distribution and both particles are cold.
Initial profile parameters are chosen as 
$f_1, f_2 \in \left\{ \sin, \cos \right\}$, $k_1, k_2 \in \left\{1,2\right\}$. Figures
\ref{fig:avgres_case1} and \ref{fig:avgres_case2} show the residual at each GMRES iteration in
the these simulation, averaged over all 30 timesteps. The thick blue lines denote residuals
observed in simulations using a plain GMRES solver and the thin, colored lines show residuals
observed when using the trained amortized for $\naug = 8, 16, 32, 48$. Using $\naug=8$ 
the average residuals of the amortized solver are only slightly smaller than those of
plain GMRES. For $\naug=16$, the average residuals of the amortized solver are of
the same magnitude as those of plain GMRES but after its following iteration ahead. That is,
the amortized solver requires one less iteration to converge to the same tolerance as plain GMRES. 
For $\naug=32$ and $\naug=48$, the amortized solver requires two fewer iterations to converge the
residual to the same tolerance as the plain GMRES.

In conclusion, we demonstrate the feasibility to accelerate plasma simulations using
implicit PIC methods with a machine-learning based amortized solver. For simulations of
electron plasma oscillations the number of required GMRES iterations per first Newton iteration
may be reduced from about 7 to about 5. The fidelity of the accelerated simulations is
unmodified and they retain all conservation properties.
Future work will focus on re-formulating the amortized solver to make use of more direct
Krylov space augmentation techniques as discussed for example in \cite{morgan-1995, chapman-1997, trivedi-2019}.
We hope to achieve a net speed-up of the simulations using such techniques. Finally, we aim to investigate 
how the performance of the amortized solver depends on the training set
and explore performance for other types of simulations such as ion acoustic waves.

\section{Acknowledgements}
The simulations presented in this article were performed on computational resources managed and supported by Princeton Research Computing,
a consortium of groups including the Princeton Institute for Computational Science and Engineering (PICSciE) and the Office of Information
Technology's High Performance Computing Center and Visualization Laboratory at Princeton University.
\bibliography{myrefs}


%

\end{document}